# Simulating Stochastic Reaction-Diffusion Systems on and within Moving Boundaries

Authors:

Atiyo Ghosh and Tatiana T. Marquez-Lago*

Integrative Systems Biology Unit, Okinawa Institute of Science and Technology, Okinawa, Japan

* Correspondence: tatiana.marquez@oist.jp

# Abstract

Chemical reactions inside cells are generally considered to happen within fixed-size compartments. Needless to say, cells and their compartments are highly dynamic. Thus, such stringent assumptions may not reflect biochemical reality, and can highly bias conclusions from simulation studies. In this work, we present an intuitive algorithm for particle-based diffusion in and on moving boundaries, for both point particles and spherical particles. We first benchmark in appropriate scenarios our proposed stochastic method against solutions of partial differential equations, and further demonstrate that moving boundaries can give rise to super-diffusive motion as well as time-inhomogeneous reaction rates. Finally, we conduct a numerical experiment representing photobleaching of diffusing fluorescent proteins in dividing *Saccharomyces cerevisiae* cells to demonstrate that moving boundaries might cause important effects neglected in previously published studies.

Keywords: Moving Boundaries, Stochastic Reaction-Diffusion, Particle Tracking

# 1. INTRODUCTION

A large number of biological systems where moving boundaries can have important effects have been identified recently. To name a few examples, it has been suggested that rapidly changing geometries in dividing cells might be important factors behind driving asymmetric cell division [1], while growing domains have already been identified as important features in models of cell migration [2] and tumour growth [3].

Previous work in the area has mainly focussed on deterministic approaches, which might prove unsuitable for many biological applications. This is especially the case when the population of reactants can be on the order of only a few individuals, or when noise may propagate within the system causing non-classic effects [4]. In a stochastic setting, there are some examples of growing domains in the context of the Reaction Diffusion Master Equation [2], where particles diffuse on a discrete lattice, and each particle is assumed to be evenly-distributed within each voxel on the lattice. However, there are significant drawbacks to the Master Equation approach: in the limit of a very fine lattice, Master Equations cease to account for bimolecular reactions [5]; and the discrete nature of the lattice fails to account for particle size, making it problematic to account for volume exclusion effects. Perhaps even more importantly for our purposes is that Master Equation approaches can introduce inaccuracies at boundaries [6] for several types of boundary conditions. Thus, there is an outstanding need for a simulation methodology incorporating both moving boundaries and stochastic effects, without the introduction of the artefacts that the Reaction Master Equation suffers from.

Alternative approaches for simulating stochastic reaction-diffusion systems with moving boundaries might draw from existing methodologies from the case of static domains. Potentially promising approaches which do not suffer from the aforementioned drawbacks treat space continuously, and can broadly be placed into two categories: simulators with fixed time-steps, and simulators with adaptive time-steps. For simulation techniques with adaptive time-steps [7, 8], the general approach is to consider imaginary domains around each diffusing particle such that the imaginary domain of one particle does not overlap with that of another. Analytical forms for the Green's functions for single particles can be used to calculate the first hitting times to each imaginary domain boundary, and to propagate the particles at each time-step chosen according to the smallest first hitting time. This allows the simulator to consider particles independently of each other at different time steps, in which case analytical solutions from single particle systems can be used. For interactions with boundaries, analytical forms for the Green's function with the appropriate boundary are needed. Extending such approaches to moving boundaries can become problematic, since Green's functions must be calculated for problems with moving boundaries, every time step. In general, these are not equations that lend themselves to analytical solutions, so numerical approaches become necessary. Schemes for handling these problems do exist, such as moving finite elements [9], or Level-Set methods [10]. These approaches are commonly used to investigate problems involving solving partial differential equations (PDEs) across phase-changes, and they can be generalized to time-varying domains. However, these numerical approaches can entail significant computational costs, which makes using them unfeasible, especially when simulating systems with many particles. It is possible to choose an adaptive time-step such that at most one particle interacts with a boundary per adaptive time-step, thus limiting the required number of PDEs with moving boundaries to be calculated to a single instance. However, this would limit the adaptive time-step to take small increments, and by consequence greatly hinder the numerical efficiency of the algorithm.

In this paper, we propose that the most promising approach for simulating reaction-diffusion in and on moving boundaries is to build on the basis of particle-tracking methods [11]. The crux of the algorithm is to choose a fixed-time step, and propagate particles according to distributions consistent with their Green's functions in free-space. If a particle should fall outside the domain after a time-step, it can be mapped back into the domain according to some rule. While such approaches have been used in static domains [11], we address here their applicability to changing domains.

We demonstrate the accuracy of our approach by numerically showing that particle-trackers can give sample paths consistent with corresponding PDEs in scenarios with moving boundaries. This investigation is conducted for point particles, as well as spherical particles with finite radius. Additionally, we investigate the effect of moving boundaries on mean-squared displacements, and how they can give rise to unexpected and non-standard profiles of reaction rates. As a final application, we numerically investigate photobleaching in a dividing *S. cerevisiae* cell. This popular experiment has so far only been investigated in static domains, and we demonstrate that moving boundaries can induce important effects that had not been accounted for.

## 2. RESULTS

### 2.1 FORMULATION

We assume the probability of a particle to simultaneously react and collide with a moving boundary to be negligibly small, thus allowing us to consider particle-boundary and particle-particle events separately. With this in mind, our aim is to construct a simulator which generates sample paths consistent with a suitable partial differential equation (PDE), i.e. the diffusion equation, for the probability density of a single particle with moving boundaries. After constructing this diffusive motion, interactions between particles may be added by established techniques.

For the following sections, we consider a closed time-dependent domain $\Omega_t \in \mathbb{R}^d$ ($d \in \mathbb{N}$) with a smooth boundary $\Gamma_t$. We denote the region outside of the boundary by $\Omega_t^c$. For every $\mathbf{x} \in \Gamma_t$, we denote the velocity and the normal of the boundary by $\mathbf{v}_t(\mathbf{x}) = \left(v_1^{(t)}, v_2^{(t)}, ..., v_d^{(t)}\right)$ and $\mathbf{n}_t(\mathbf{x}) = \left(n_1^{(t)}, n_2^{(t)}, ..., n_d^{(t)}\right)$, respectively. Note that we consider a *d*-dimensional Euclidean space so as to provide a general setting for our algorithm.

## 2.1.1 Diffusion and reaction events within moving boundaries

We assume that the probability density function (PDF) of a single particle in $\Omega_t$ undergoing Brownian motion obeys the following relation:

$$\frac{\partial P}{\partial t} = D\nabla^2 P \text{ for } \mathbf{x} \in \Omega_t \text{ with } \nabla P = -(\mathbf{v}_t \cdot \mathbf{n}_t)P \text{ on } \Gamma_t, \quad (1)$$

where the boundary condition has been derived by imposing the conservation of probability. Namely, assuming that the boundary condition takes the form $\nabla P = g(P, \mathbf{v}_t, \mathbf{n}_t)$ for all $\mathbf{x} \in \Gamma_t$, the following must hold:

$$\begin{aligned}\frac{d}{dt}\int_{\Omega_t} P d\mathbf{x} &= \int_{\Omega_t} \frac{\partial P}{\partial t} d\mathbf{x} + \int_{\Gamma_t} (\mathbf{v}_t \cdot \mathbf{n}_t) P d\mathbf{x} \\ &= \int_{\Gamma_t} \left(g(P,\mathbf{v}_t,\mathbf{n}_t) + (\mathbf{v}_t \cdot \mathbf{n}_t)P\right) d\mathbf{x} = 0\end{aligned}, \quad (2)$$

where the first equality follows by applying Leibniz's integral rule. The second equality follows directly from the Divergence Theorem, namely by equating the rate of change of probability in the domain with the flux of probability through the domain boundary. Comparing terms yields the boundary condition in (1). Note that for a static domain, $\mathbf{v}_t(\mathbf{x}) = 0$ for all $\mathbf{x} \in \Gamma_t$ holds, and the boundary condition takes the form of a zero-Neumann type. Note that we concentrate here on Neumann boundary conditions. However, generalisations of the particle tracker to incorporate effects of Robin boundary-conditions directly follow from existing methods of semi-permeable membranes in the case of static boundaries [11].

To simulate trajectories from (1), we constructed an algorithm represented by the pseudo-code in Table (1) and graphically in Fig. 1a. In summary, we introduce $n \in \mathbb{N}$ particles, which we enumerate with some index $i \in \{0, 1, \dots, n\}$, each with radius $\rho_i$ and associated diffusion coefficient $D_i$. Note that the explicit incorporation of particle size allows this method to circumvent problems associated with Master Equations (please refer to the Introduction for further details).

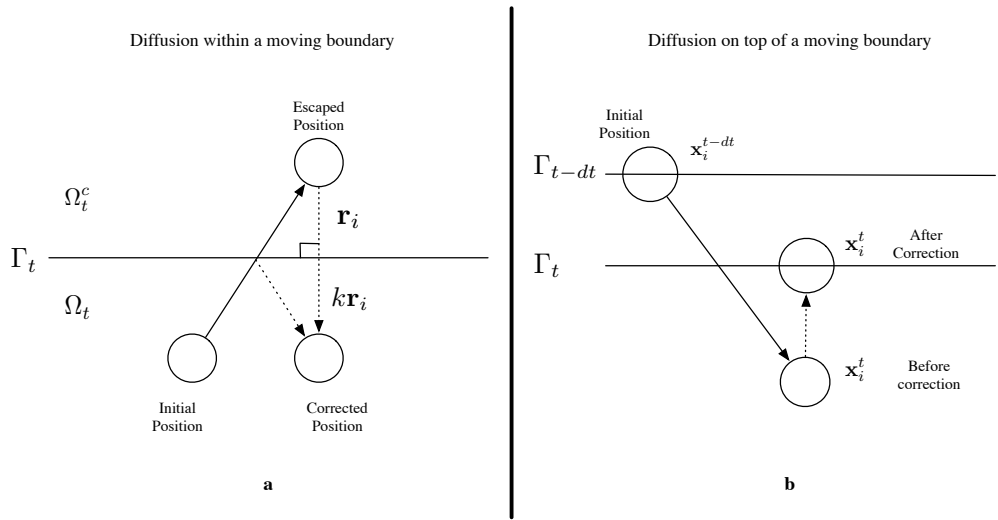

Figure 1: Schematics of proposed algorithms. a) A visual representation of the proposed method in Table 1, mapping escaped particles back into the domain

for the diffusion within moving boundaries. b) A visual representation of the proposed method in Table 2 mapping diffusing particles on a moving domain.

Each particle is propagated in a small time step, $dt$, along each Cartesian coordinate according to a Normal distribution with mean zero and variance $2D_i dt$. At each time step $t$, we check for any particles lying outside $\Omega_t$ or within the particle radius away from $\Gamma_t$. For each of these particles, we calculate the point on $\Gamma_t$ which is closest to the escaped particle, and construct a vector, $\mathbf{r}$, from the particles position, $\mathbf{x}_i$, to this point. For point particles (i.e. $\rho_i = 0$), we set $\mathbf{x}_i^{(new)} = \mathbf{x}_i - k\mathbf{r}_i$ for some $k > 1$. For particles with positive radius, we place the particles at the closest point on the domain, and then move them in the normal direction by $\rho_i$. This ensures that the particle is contained within the domain. Given $n \in \mathbb{N}$ particles, define their positions $\mathbf{x}_i^{(0)} = \left( x_{i1}^{(0)}, x_{i2}^{(0)}, \ldots, x_{id}^{(0)} \right) \in \mathbb{R}^d$ for $i = 1, 2, \ldots, n$ such that $\mathbf{x}_i^{(0)} \in \Omega_0$ and $\left| \mathbf{x}_i^{(0)} - \mathbf{y}_0 \right| > \rho$ for all $\mathbf{y}_0 \in \Gamma_0$. Define a maximum time of the simulation, $t_{max}$. Set $t = 0$ and define some small $dt$.

---

While $t < t_{max}$:
   $t = t + dt$
   For $i = 1, \ldots, n$:
      $\mathbf{x}_i^{(t)} = \left( x_{i1}^{(t-dt)} + N(0, 2D_i dt), x_{i2}^{(t-dt)} + N(0, 2D_i dt), \ldots, x_{id}^{(t-dt)} + N(0, 2D_i dt) \right)$
   For all $i$ such that $\left| \mathbf{x}_i^{(0)} - \mathbf{y}_t \right| < \rho_i$ with $\mathbf{y}_t \notin \Gamma_t$ or $\mathbf{x}_i^{(t)} \notin \Omega_t$:
      Find $\mathbf{y}_i \in \Gamma_t$ which minimizes $\left| \mathbf{x}_i - \mathbf{y}_i \right|$
      Set $\mathbf{r}_i = \mathbf{x}_i - \mathbf{y}_i$
      Set $\mathbf{x}_i = \mathbf{x}_i - k\mathbf{r}_i$ if $\rho = 0$
      Set $\mathbf{x}_i = \mathbf{x}_i - k\mathbf{r}_i$ then move particle in the direction of $\mathbf{n}$ by $\rho$ if $\rho > 0$.
   For all $i = 1, \ldots, n$:
      For all $j = 1, \ldots, n$ and $j \neq i$:
         If $\left| \mathbf{x}_i - \mathbf{x}_j \right| < \rho_{ij}^{(reac)}$:
            Replace particles $i$ and $j$ with reactants.
end While

---

Table 1: Pseudo-code representing the proposed algorithm for simulating reaction-diffusion systems inside the volume confined by a moving boundary.

Regarding chemical reactions, we execute a loop over all pairs of particles to calculate the distance between each possible pair of particles. If reactive particles $i$ and $j$ are separated by a distance less than or equal to $\lambda_{ij}^{(r)}$, we execute reaction $r$ accordingly. In case there is a single reaction product, we choose the location of the product to be at the same position (unary reactions) or in the middle position between the two reactants (binary reactions). In case there are two products, we place the particles in the same locations as the original reactants for the case of binary reactions, or at randomly opposite poles of a small sphere centred on the reactant in the case of unary reactions. The size of the sphere is chosen to be of the order of the largest reaction radius, $\lambda_{ij}^{(r)}$. In some scenarios, this might cause artefacts (for instance, in case of 'germinate' reactions, where reaction products react with each other immediately). In these cases, and in cases for a higher number of reaction products, alternative procedures from static boundary scenarios can be used [11]. For the case that two particles' radii overlap with each other after a time step, the particles are then moved an equal amount away from each other in the direction defined by the vector connecting the particles' centres.

Note that the algorithm represented above allows for a variety of effects at the boundary. For example, setting $k=1$ amounts each escaped particle to be placed at the closest point on the domain boundary, and using $k=2$ amounts to reflecting a point particle through the closest point on the domain boundary to a point within the domain boundary. We seek to demonstrate first that, for point particles without reactions, the sample paths generated from our algorithm are consistent with (1) for a variety of different boundary effects. We provide a numerical investigation of this in section 2.2.1.

### *2.1.2 Diffusion and reactions on a moving boundary*

We assume that the PDF for a diffusing species on $\Gamma_t$ obeys the following equation:

$$\frac{\partial P}{\partial t} = D\nabla^2_{\Gamma_t} P - \mathbf{v}_t \nabla P \qquad (3)$$

where $\nabla^2_{\Gamma_t} = (\nabla - \mathbf{n}_t(\mathbf{n}_t.\nabla)).(\nabla - \mathbf{n}_t(\mathbf{n}_t.\nabla))$ refers to the surface Laplacian. Note that if $P$ is initially defined on $\Gamma_0$, then the advection term in (3) will act in a way such that $P$ remains on $\Gamma_t$ for all $t$. To simulate sample paths consistent with (3), we use the algorithm represented by the pseudo-code in Table 2, and graphically in Fig. 1b. As before, we have $n \in \mathbb{N}$ particles with the same index set $i \in \{0,1,...,n\}$, with particle radii $\rho_i$ and diffusion coefficients $D_i$. Particles $i$ and $j$ react in a similar way as the previous section, i.e. when their centres are separated by a distance of less than $\lambda_{ij}^{(r)}$. At every time-step, each particle is propagated in a small time step, $dt$, along each Cartesian coordinate according to a Normal distribution with mean zero and variance $2D_i dt$. For every time step from $t$ to $t+dt$, the particles are further propagated by the amount $\mathbf{v}_t dt$. For each particle, we then calculate the point on $\Gamma_t$ which is closest to the escaped particle, and map each particle onto this point. Then, we execute reactions and collisions exactly as per section 2.1.1. Note that this procedure is independent of the curvature of the surface of the boundary. A heuristic justification for this follows from considering the limit of small time steps, such that the diffusive motion is small, and the surface is locally flat. Since the projection of a Brownian motion itself is Brownian, we expect that this procedure should give consistent sample paths to (3).

> Given $n \in \mathbb{N}$ particles, define their positions $\mathbf{x}_i^{(0)} = \left(x_{i1}^{(0)}, x_{i2}^{(0)}, ..., x_{id}^{(0)}\right) \in \mathbb{R}^d$ for
> $i = 1, 2, ..., n$ such that $\mathbf{x}_i^{(0)} \in \Gamma_0$. Define a maximum time of the simulation, $t_{max}$.
> Set $t = 0$ and define some small $dt$.
> While $t < t_{max}$:
>   $t = t + dt$
>   For $i = 1, ..., n$:
> $\mathbf{x}_i^{(t)} = \left(x_{i1}^{(t-dt)} + N(0, 2D_i dt) + v_1^{(t-dt)} dt, x_{i2}^{(t-dt)} + N(0, 2D_i dt) + v_2^{(t-dt)} dt, ..., x_{id}^{(t-dt)} + N(0, 2D_i dt) + \right.$
>
>   Find $\mathbf{y}_i \in \Gamma_t$ which minimizes $|\mathbf{x}_i - \mathbf{y}_i|$
>   Set $\mathbf{x}_i^{(t)} = \mathbf{y}_i$
>   For all $i = 1, ..., n$:
>     For all $j = 1, ..., n$ and $j \neq i$:
>       If $|\mathbf{x}_i - \mathbf{x}_j| < \rho_{ij}^{(reac)}$:
>         Replace particles $i$ and $j$ with reactants.
> end While

Table 2: Pseudo-code representing the proposed algorithm for simulating reaction-diffusion systems occuring on the surface of a moving boundary.

## 2.2 ALGORITHM BENCHMARKING AND SAMPLE THEORETICAL APPLICATIONS

### 2.2.1 Diffusion within and on a shrinking circle

We first consider diffusion within and on a shrinking circle as a sample application. We first demonstrate that the sampled particle tracks generated with our algorithm are consistent with the solutions of the PDEs in (1) and (3), then proceed to investigate the effects of moving boundaries on the mean squared displacement (MSD) of diffusing particles. Note that we treat PDEs as descriptions of the PDFs of individual particles, as opposed to the behaviour of diffusion of many particles. Consider $\Omega_t$ to be the area bound by a circle in $\mathbb{R}^2$, cantered at the origin, with $\Gamma_t$ denoting its boundary. At time $t$, we denote the radius of such a circle by $r_t$. The starting radius is written $r_s$, and the finishing radius as $r_e$. Simulations occur up to an end time of $t_e$, and the change of $r_t$ from the starting radius to the end is taken to be linear in time. We denote the Euclidean distance of a particle at time $t$ from its initial position by $x_t$. For $x_t \propto t^\alpha$, we denote the process as anomalous diffusion: of the sub-diffusive type for $\alpha < 1$ and super-diffusive for $\alpha > 1$. Diffusion is denoted as Brownian, or non-anomalous, if $\alpha = 1$.

## 2.2.1.1 Diffusion inside the volume of a shrinking circle

We numerically demonstrate convergence by considering the cumulative density functions (CDFs) $P(R < r_t)$ in shrinking circle domains, incorporating diffusion constants in widely different scales. Figure 2 shows results considering several scenarios: 1) the diffusive motion of each particle is much faster than that of the boundary (Figure 2a and 2d), 2) particles and the boundary moving at a similar rates (Figure 2b and 2e) and, 3) particles diffusing much slower than the boundary (Figure 2c and 2f). We investigate these three scenarios for both point particles (Figure 2a-c) and for particles with finite size (Figure 2d-f). For point particles, we consider two methods by which to map escaped particles back into $\Omega_t$: we place escaped particles at the closest point on the boundary (i.e. $k=1$ in the pseudo-code in Table 1), or we reflect the particles through the closest point in the boundary (i.e. $k=2$). Solutions to (1) were found adopting a deforming mesh from the commercial software package Comsol Multiphysics version 4.3. To allow for particle size to be included into the PDE solutions, the value $r_t$ was reduced by an amount $\rho$ for all $t$. Stochastic simulations were run with single particles with initially uniformly distributed positions, and repeated 2000 times. Please refer to the legend of Figure 2 for the time-step of stochastic simulations. In these scenarios, we find that the CDFs generated by the particle tracker and PDEs are in good agreement, and this is independent of the correction method used to map escaped particles.

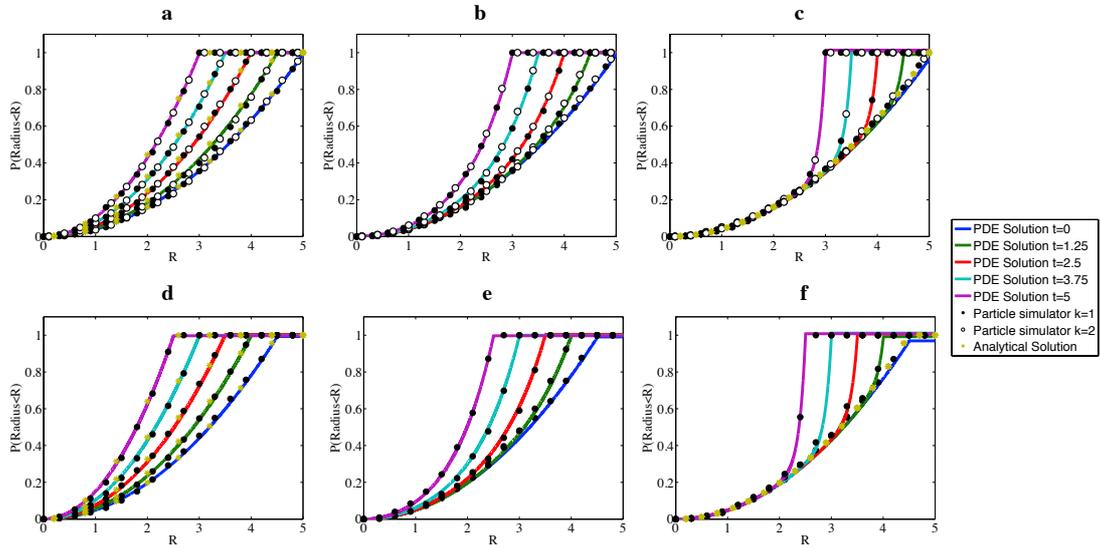

Figure 2: Cumulative density functions obtained from numerical and analytical methods. The CDF of particles diffusing within a shrinking circle, where the radius shrinks from an initial value of 5 length units to a final radius of 3 length units uniformly over a 5 second interval. The CDF is plotted at times 0 (purple), 1.25 (light blue), 2.5 (red), 3.75 (green) and 5 (blue) seconds.

Lines represent the solutions as calculated from PDEs. Black dots represent results from the algorithm of Table 1 with *k*=1, hollow dots represent the same except with *k*=2. Yellow stars dots represent the predictions on the basis of a uniform distribution. Subplots a to c represent point particles with diffusion constants 5, 0.5 and 0.05 unit squared per second respectively, and subplots d to f represent particles of radius of 0.5 units, with diffusion constants 5, 0.5 and 0.05 unit squared per second, respectively. Time steps for simulations are plots are 0.001s (for *D*=0.05), 0.0001s (for *D*=0.5) and 0.00001 (for *D*=5). Simulations were repeated 1500 times

In cases where the diffusion is much faster than the boundary movement, we observe a uniform distribution in the circle at all times (i.e. $P(R<r_t)=R^2/(r_t-\rho)^2$). In situations where diffusive motion is significantly slower than the boundary movement we observe that the CDFs of particles remain consistent with the uniform distribution over the initial circle with radius $r_s$, rising quickly to unity close to the boundary of the sphere i.e.:

$$P(R<r_t) = \begin{cases} (R-\rho)^d/(r_s-\rho)^d & \text{for } R < r_t - \rho \\ 1 & \text{for } R = r_t - \rho \end{cases}. \qquad (4)$$

These analytical results are in good agreement with numerical experiments, as shown in Figure 2c and 2f.

This model system also allows us to consider the effects of moving boundaries on the mean squared displacement of diffusing particles. For doing this, we denote the Euclidean distance of a particle at time *t* from its initial position by $x_t$. For large diffusion coefficients, we observe that the mean-squared displacement is consistent with freely diffusing particles ($\langle x_t^2 \rangle \propto t$) so long as we consider short time scales to reduce the effects of diffusion within confined volumes (note the coefficient close to unity of the linear regression represented by red dots in Figure 3c). At larger times, the beginning of confinement effects give rise to sub-diffusive behaviour (note the black dots in Figure 3c). Assuming the particles are diffusing very slowly relative to the boundary, we conjecture that the MSD can be decomposed into two components: the fraction of freely diffusing particles, and those being pushed by the boundary. Enumerating these terms results in the following equation, where the first term represents freely diffusing particles, and the second term represents those being swept from their initial positions by the boundary:

$$\begin{aligned}\langle x_t^2 \rangle &= \frac{4r_t^2}{r_s^2}Dt + \int_{r_t}^{r_s} \frac{2\pi r}{\pi r_s^2}(r-r_t)^2 \, dr \\ &= \frac{4r_t^2}{r_s^2}Dt + \frac{2}{r_s^2}\left(\frac{1}{4}(r_s^4-r_t^4) - \frac{2}{3}r_t(r_s^3-r_t^3) + \frac{r_t^2}{2}(r_s^2-r_t^2)\right)\end{aligned}. \qquad (5)$$

In more detail: The first term represents the product of the proportion of freely diffusing particles ($\frac{r_t^2}{r_s^2}$) with the MSD of freely diffusing particles in free space ($4Dt$); the second term is an integral of the proportion of particles close to the boundary, integrated over the distance the boundary moves.

Remembering that $r_t$ is changing linearly in time, we thus expect to observe super-diffusive behaviour at small diffusion constants. These findings are in good agreement with our simulations. Please see Figure 3a for details.

Note that such arguments so far would neglect any effect of growing domains, i.e. where the boundary would be moving away from particles. In such scenarios, we would expect our current methodology to give non-anomalous MSD characteristics. However, it may well be that such boundaries give rise to an advective velocity on moving particles and might still provide interesting effects. In such scenarios, advective forces can be introduced into particle tracking methodologies, but we consider this is outside the scope of the current paper.

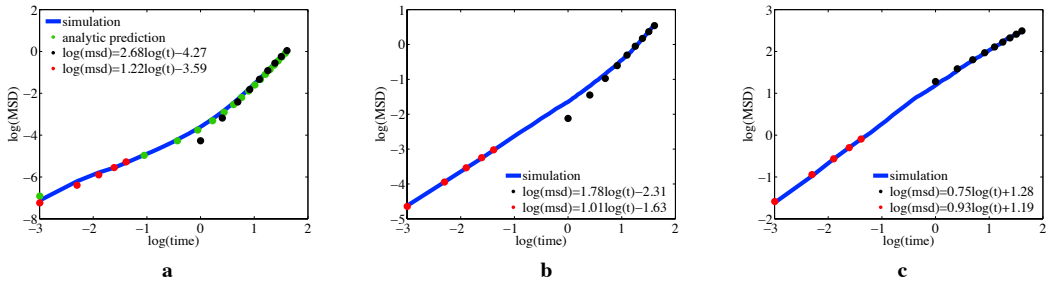

Figure 3: Mean-squared displacement for diffusion inside a shrinking sphere. Log-Log plots of the MSD against time for particles diffusing within a shrinking sphere, where the radius shrinking from an initial value of 5 length units to a final radius of 3 length units uniformly over a 5 second interval, with subplots a-c showing results for diffusion constants of 0.005, 0.05 and 1 respectively. Blue lines show simulation results of single particles on the basis of 2000 repeats. Red dots show a linear regression on the basis of data from the first two seconds, black dots show linear regression on the basis of the last second. Green dots show the predictions of equation (5).

## *2.2.2 Diffusion on the surface of a shrinking circle*

We now consider the case of diffusion on the surface of a shrinking circle. In this scenario, numerical solutions of (3) can be found by considering the problem in polar coordinates. We perform the transformation in two steps, first by considering a transformation to arc-lengths along the circle (which removes advective terms due to the moving boundary), then to arc-angles (which transforms the problem from a moving boundary to a static boundary problem). We first define the position of a particle in terms of $\tilde{r}_t \in (-\pi r_t, \pi r_t]$, which is the arc-distance along the shrinking circle of the particle from the point where the circle intersects the positive *y*-axis. We take $\tilde{r}_t$ to be positive (negative) for coordinates with a positive (negative) *x* coordinate, to denote the specific displacement. No advective term results in the transformed diffusion equation, since the movement of the surface is perpendicular to the diffusive motion along surface. Now we can reformulate the problem in terms of a diffusion equation for the PDF $\bar{P}(\tilde{r}_t, t)$ of the particle being at $\tilde{r}_t$ at time $t$:

$$\frac{\partial \bar{P}}{\partial t} = D \frac{\partial \bar{P}}{\partial \tilde{r}_t^2} \quad \text{with} \quad \bar{P}(-\pi r_t, t) = \bar{P}(\pi r_t, t) . \tag{6}$$

The above equation is a moving boundary problem, but can be reformulated into a static domain by a change of variables to the angle from the positive $y$ axis, here denoted by $\theta \in (-\pi, \pi]$, and we write the corresponding PDF as $\hat{P}(\theta, t)$. Using the relationship $\tilde{r}_t = r_t \theta$, we can transform (6) into the following equation.

$$\frac{\partial \hat{P}}{\partial t} = \frac{D}{r_t^2} \frac{\partial^2 \hat{P}}{\partial \theta^2} \quad \text{with} \quad \hat{P}(-\pi, t) = \hat{P}(\pi, t) , \tag{7}$$

which is a diffusion equation with a time-varying diffusion coefficient, but with periodic non-moving boundaries. This formulation lends itself to a simple numerical solution via an Euler method, and allows us to benchmark the proposed algorithm. Numerical investigations over a range of diffusion coefficients show that our algorithm generates consistent sample paths with (7), as shown in Figure 4a-c.

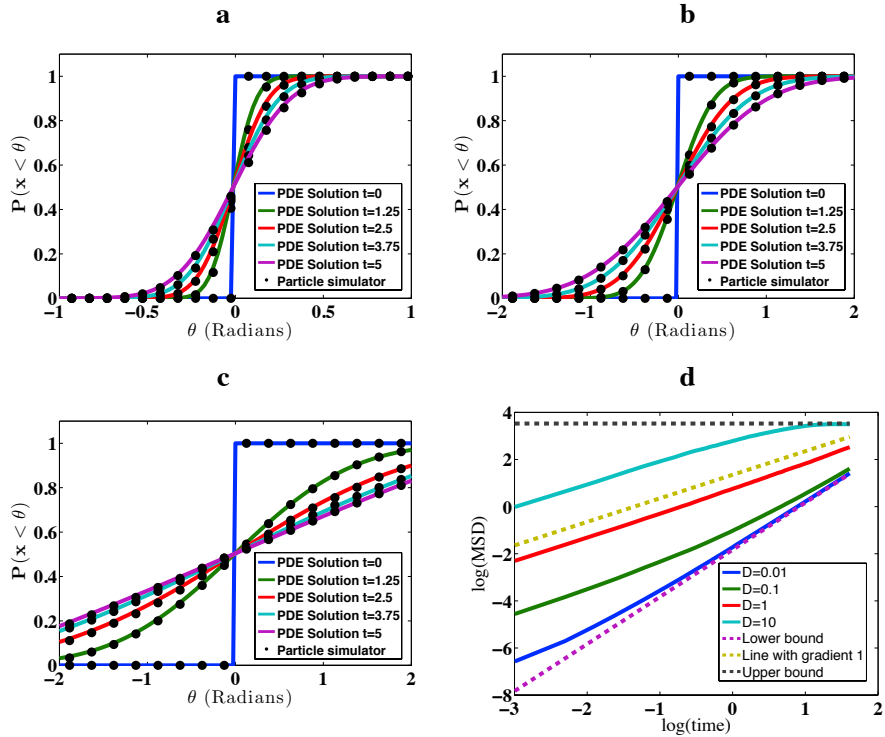

Figure 4: Diffusion on the surface of a shrinking circle. Subplots a-c: CDFs of the angular component of particles diffusing on the surface of shrinking sphere, where the radius shrinks from an initial value of 5 length units to a final radius of 3 length units uniformly over a 5 second interval at times 0 (blue), 1.25 (green), 2.5 (red), 3.75 (light blue) and 5 (purple) seconds. Lines show numerical results from PDEs, dots show results from simulations of single particles with 10000 repeats. Diffusion constants are 0.1 (subplot a), 1 (subplot b) and 10 (subplot c) units squared per second, respectively. Time steps for simulations are plots are 0.0005s (for $D=0.1$), 0.00005s (for $D=1$)

and 0.000005 (for *D*=10). Subplot d) shows MSD curves for the same simulations from subplots a-c in green, red and light blue, respectively. In addition, the blue line at the bottom represents results with a diffusion constant of 0.01 length units per second squared. Also shown are predictions of equations (8) and (9) (dotted black and dotted purple line respectively), and a dotted yellow line with gradient 1. The latter represents a non-anomalous regime to aid judgement of where diffusion is super or sub-diffusive.

We further investigated the effect of the moving boundary on the MSD. At large diffusion coefficients, the simulated behaviour is consistent with free diffusion at the start, before becoming sub-diffusive as the MSD approaches an upper bound (due to the finite circumference of the circle at the end of the process). To derive an expression for the upper bound of the MSD, assume that the distribution of particles is uniform over the circle at the end time. Considering the initial particle position of $(0, r_s)$, and a final position $(r_e \cos\theta, r_e \sin\theta)$ which is uniformly distributed over $\theta \in [0, 2\pi)$, this gives the following expression:

$$\langle x_{t_e}^2 \rangle = \frac{1}{2\pi} \int_0^{2\pi} \left( (r_s - r_e \cos\theta)^2 + r_e^2 \sin^2\theta \right) d\theta$$
$$= r_s^2 + r_e^2$$
(8)

In the limit of zero diffusion rates, we can derive the limit behaviour by assuming the diffusing particle is static on the moving domain at the point $(0, r_t)$ for all *t*. This trivially gives the expression:

$$\langle x_t^2 \rangle = (r_s - r_t)^2 .$$
(9)

Recalling that $r_t$ changes linearly in time, this implies that the motion is super-diffusive if the displacement due to the moving boundary is larger than that related to particle diffusion. As diffusion rates increase such that the motion due to diffusion is of comparable to that of boundary movement, we observe that super-diffusive effects subside, as shown in Figure 4d. Now that we have presented our algorithm in full, we will focus on sample applications to illustrate its ease of use and applicability.

## 2.3 SAMPLE EXPERIMENTAL APPLICATIONS

### *2.3.1 Bimolecular decay inside an elongating dumbbell*

We now consider some non-standard effects that moving boundaries might have on reaction kinetics. The domain in consideration is two spheres joined together by a bridge, representing an elongating dumbbell. This shape resembles dividing nuclei in closed mitosis [12], among other biophysical processes. Initially, the bridge is of length zero. After some time, the length of the bridge increases steadily toward a final value, upon which it stops growing. A schematic of the final shape of the sphere projected in 2D is found in Figure 5. Thus, we have a domain in $\mathbb{R}^3$ with rotational symmetry around the *x*-axis. We consider a particle to be within the domain so long as the position of each particle $(x(t), y(t), z(t))$ satisfies:

$$\begin{cases} (x+R-o)^2 + y^2 + z^2 < R^3 & \text{for } -2R+o < x < -l(t)/2 \\ y^2 + z^2 < d^2 & \text{for } -l(t)/2 < x < l(t)/2 \\ (x+R-o)^2 + y^2 + z^2 < R^3 & \text{for } l(t)/2 < x < 2R-o \end{cases} \quad (10)$$

for $R \in \mathbb{R}$ with $o = R - \sqrt{R^2 + d^2}$. This system represents two spheres of radius $R$, joined together by a cylinder of diameter $d$ with length $l(t)$, a geometry common in dividing cells/nuclei. The centre of the cylinder is at $x=0$.

We then consider bimolecular decay $A + B \rightarrow \varnothing$, where particles of species A and B react when they are within a distance $\lambda$ at the end of a time step. In this example, the probability of reacting is always 1 whenever particles are within that predefined distance. However, the algorithm can be generally set up in such a way that the reaction probabilities take values in the interval [0,1] once a distance threshold is reached.

As an initial condition, we place a single A molecule at the centre of the left sphere, and a single B molecule at the centre of the right sphere. We then define the length function $l(t)$ as follows:

$$l(t) = \begin{cases} 0 & \text{for } t < T_1 \\ L(t - T_1) & \text{for } T_1 \leq t < T_2 \\ L(T_2 - T_1) & \text{for } t \geq T_2 \end{cases} \quad (11)$$

i.e. the length is monotonically non-decreasing. In such scenarios, even though the total volume of the system is increasing, we note that the time-distribution of the decay event can be bimodal, as shown in Figure 5. This interestingly implies the effective reaction rate can increase with increasing volumes.

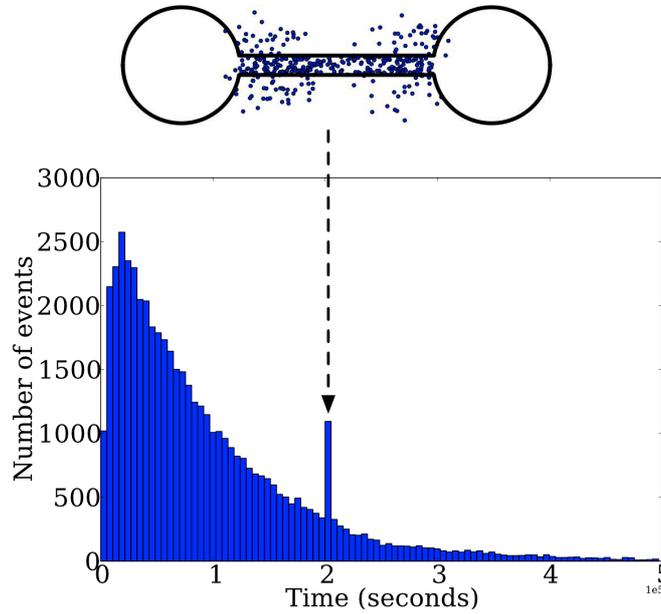

Figure 5: Bimodal reaction times in a growing domain. A histogram showing bimodal reaction times in a growing volume of two spheres connected by an elongating bridge as per section 2.2.3, with $R=3$, $d=0.5$, $L=0.01$ per second, $T_1 = 200000$ seconds, $T_2 = 201000$ seconds, $\lambda = 0.2$. Each particle has a diffusion rate of 0.0025 units squared per second, and simulation time steps are in increments of 1 second. A projection of the geometry at $T_2$ onto the x-y plane is shown, superimposed onto the location of all reaction events

occurring during bridge elongation. The location of a reaction event is taken as the mean position of the two reactants.

In Figure 5, we allow the bridge to remain at length zero for 200,000 seconds, before allowing the bridge to grow uniformly to reach length 10 units at 201,000 seconds. The spheres have a radius of 3 units at all times. Each particle diffuses slowly, with a rate of 0.0025 units squared per second. Particles react when they reach a distance of less than 0.4 units of each other. The histogram was constructed from 50,000 events, with the bin-size determined by the Freedman-Diaconis rule [13]. As the bridge elongates, we observe an increased frequency of reaction events inside the bridge. For illustration purposes, Figure 5 shows a superposition of the geometry (projected onto the *x-y* plane) at 201,000 seconds on all reaction events occurring between 200,000 and 201,000 seconds. Note that only the final geometry is depicted in Figure 5. By consequence, some reaction events occurring before the geometry reaches its final conformation appear outside the domain.

### 2.3.2 Photobleaching in and on a dividing budding yeast nucleus

As a final application, we consider photobleaching experiments done in a dividing *S. cerevisiae* nucleus during anaphase. Note that *S. cerevisiae* undergoes closed mitosis, whereby the nucleus remains intact throughout anaphase, only to break down moments before cytokinesis. Our numerical simulations aim to reproduce a more physically realistic scenario of *S. cerevisiae* nuclear division than those considered in previously studies [12, 14].

We represent mother and daughter nuclear lobes by prolate ellipsoids connected by a cylindrical bridge, as depicted in Figure 6. The major axis of the mother and daughter, as well as the axis of the cylinder, are taken to be co-axial to the *x*-axis. Initially, we construct the dividing nucleus in Cartesian coordinates $(x',y',z') \in \mathbb{R}^3$, with the extreme edge of the mother lobe coinciding with *x'*=0, followed by the bridge, followed by the daughter lobe. Then we transform the coordinate system such that the centre of mass of the nucleus lies in the centre of a coordinate system $(x,y,z) \in \mathbb{R}^3$. We denote the major and minor axes of the mother and daughter by $2r_{xm,t}, 2r_{ym,t}$ and $2r_{xd,t}, 2r_{yd,t}$ respectively. We take the ellipsoid to be rotationally symmetric around the *x*-axis. Furthermore, we define the length of the bridge to be $l_t$ and its width (i.e. radius) to be $w_t$.

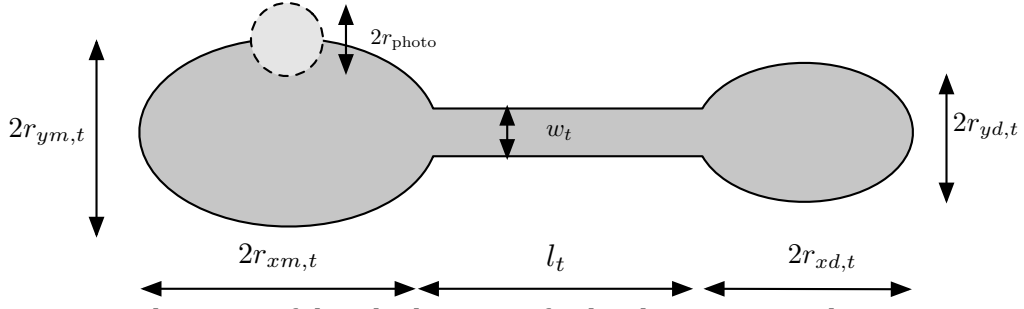

Figure 6: A schematic of the idealisation of a dividing yeast nucleus. Parameters $r_{ym,t}$ and $r_{xm,t}$ represent the maximal radius of the mother nuclear lobe in the direction of the *y* and *x* axis respectively at time *t*. We define $r_{yd,t}$ and $r_{xd,t}$ analogously for the bud nuclear lobe. The length and width of the bridge is given by $l_t$ and $w_t$ respectively, and the radius of the spherical photobleaching spot is given by $r_{photo}$. Note that the geometry is 3D, and only a 2D projection is depicted here.

Separately, images of dividing nuclei were obtained by confocal microscopy on a Zeiss LSM-780 microscope, courtesy of Dr. Jai Denton. The strain visualized, GA-1320 containing the pAA4+lacO plasmid, was kindly provided by Professor Susan Gasser and has been previously described in [1]. For imaging, live *S. cerevisiae* cells were grown overnight in 5ml SC-his-ura medium, diluted 400ul in 5ml of fresh SC-his-ura medium and grown for a further 4 hours before being harvested and immobilised between a 3% low melting temperature agarose gelpad and a coverslip for visualisation. The confocal images and the idealised geometry constructed from them can be found in Figure 7.

Geometrical parameters were measured manually with Image J from these images with time intervals of 50 seconds between successive frames. For intermediate times between frames, values were linearly interpolated to allow the shape to change smoothly (see sample videos of simulations in the Supplementary Online Material). Each ellipsoid is truncated such that the aperture of the bridge and the ellipsoid coincide without any gaps. Thus, particles in the mother cell have an *x*-component which satisfies $0 \leq x' < r_{xm,t} + r_{xm,t} \cos\left(\arcsin\left(\frac{w_t}{r_{ym,t}}\right)\right)$, with the arcsin being taken at the smallest positive value. The set of all points in the mother lobe can be then described as:

$$(x',y',z') = \left(r_{x,t}\cos\theta + r_{xm,t}, r_{y,t}\sin\theta\sin\phi, r_{y,t}\sin\theta\cos\phi\right), \qquad (12)$$

with $r_{x,t} \in [0, r_{xm,t}]$, $r_{y,t} \in [0, r_{ym,t}]$, $\phi \in [0, 2\pi)$ and $\theta \in \left[\arcsin\left(\frac{w_t}{r_{ym,t}}\right), \pi\right]$. Note that the limits for *x'* in the mother lobe stop short of $2r_{ym,t}$, since a small amount of the ellipsoid is truncated so as to attach the bridge. The area in the bridge consists of

those points for which the following relationship holds
$r_{xm,t} + r_{xm,t} \cos\left(\arcsin\left(\frac{w_t}{r_{ym,t}}\right)\right) \leq x' < r_{xm,t} + r_{xm,t} \cos\left(\arcsin\left(\frac{w_t}{r_{ym,t}}\right)\right) + l_t$:

$$(x',y',z') = (x', w_t \sin\phi, w_t \cos\phi) \qquad (13)$$

with $\phi \in [0, 2\pi)$. Finally, the points in the daughter lobe satisfy
$r_{xm,t} + r_{xm,t} \cos\left(\arcsin\left(\frac{w_t}{r_{ym,t}}\right)\right) + l_t \leq x' < r_{xm,t} + r_{xm,t} \cos\left(\arcsin\left(\frac{w_t}{r_{ym,t}}\right)\right) + l_t + r_{xd,t} + r_{xd,t} \cos\left(\arcsin\left(\frac{w_t}{r_{yd,t}}\right)\right)$,

where the largest value less than $\pi$ is taken from the arcsin and:

$$(x',y',z') = \left( r_{xm,t} + r_{xm,t} \cos\left(\arcsin\left(\frac{w_t}{r_{ym,t}}\right)\right) + l_t + r_{xd,t} + r_{x,t} \cos\theta, r_{y,t} \sin\theta \sin\phi, r_{y,t} \sin\theta \cos\phi \right) \qquad (14)$$

for $r_{x,t} \in [0, r_{xd,t}]$, $r_{y,t} \in [0, r_{yd,t}]$, $\phi \in [0, 2\pi)$ and $\theta \in \left[0, \arcsin\left(\frac{w_t}{r_{yd,t}}\right)\right]$.

This geometry is then translated parallel to the *x*-axis such that the centre of mass is situated at the origin for all times. The geometry is updated at every time step of the simulation, ensuring a smoothly changing geometry and thus minimising numerical artefacts that might arise from more sudden changes of geometry. The velocity field of the moving boundary is calculated numerically, and is assumed to act in a way such that the velocity carries each point on a membrane to its nearest point at the subsequent time. This assumption might not be true for dividing *S. cerevisiae* nuclei, but such data is difficult to measure empirically, and so here we must resort to some assumption of this nature.

Finally, we consider a 'bleaching' region in which a degradation reaction $A \xrightarrow{k_{deg}} \emptyset$ occurs, where $k_{deg}$ gives the rate per unit time that a single protein *A* degrades due to photobleaching. The region where photobleaching can occur is taken to be the volume of a sphere of radius $r_{photo}$ centred at $(r_{xm,t}, r_{ym,t}, 0)$. That is, on the edge of the centre of the mother lobe, and such region will be denoted as the 'bleach spot' to conform with terminology used in [12, 14]. At the end of each time step of length *dt*, any particles lying inside the photobleaching region can decay according to a Bernoulli trial with probability $k_{deg} dt$.

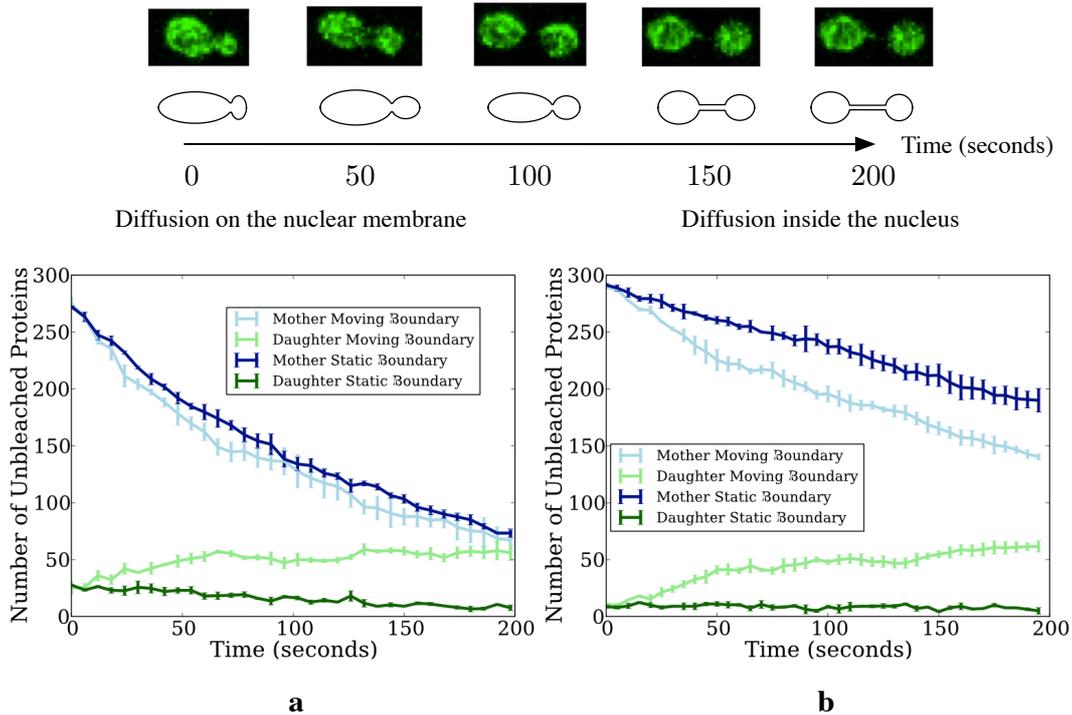

Figure 7: Photobleaching in a dividing yeast nucleus. Plots showing qualitatively different behaviour between static and moving boundaries for diffusion on (subplot a) and within (subplot b) a dividing yeast nucleus. The population of particles in the mother and daughter lobes are in blue and green respectively, with dark hues and light hues representing the static and moving boundary cases, respectively. Diffusion coefficients for all plots are 0.25 $\mu m^2/s$, with $r_{photo} = 0.17 \mu m$ and $k_{deg} = 2$. The change of geometries is depicted in the diagram above the plots, with accompanying images from confocal microscopy. Time steps for all simulations are 0.00005 seconds.

To compare the effect of moving boundaries, we conducted two sets of simulations: one set with static boundaries, and another where boundaries were allowed to change with time. These sets of simulations were repeated for proteins diffusing both in the membrane and on the membrane. We define the middle of the bridge to mark the point where proteins switch between mother and daughter lobes. Plots for the number of *A* proteins in the mother and daughter lobes are shown in Figure 7. These are split between diffusion on the nuclear membrane (Figure 7a) and diffusion inside the nucleus (Figure 7b). Three runs of simulations with 300 proteins per run were conducted, with proteins initially uniformly distributed through the whole nucleus. In the case of diffusion on the nuclear membrane, we find that the population of *A* proteins in the mother lobe stays similar between the static and the growing cases. However, the growth of the domain allows a significant proportion of proteins to avoid photobleaching by diffusing/escaping to the daughter lobe. We observe that the average number of unphotobleached proteins on the entire nuclear membrane are 80 and 122 for the case of static and moving boundaries respectively. Thus, the effective photobleaching rate for scenarios for diffusion in this static scenario is higher than that of the moving boundaries scenario. Thus, neglecting this effect might lead to underestimates of the bleaching rate in FLIP and FRAP studies. For photobleaching experiments performed for particles diffusing inside the nucleus, we observe a similar overall photobleaching rate between the static and moving cases. However, the distribution of proteins is markedly different: decay in the mother lobe is significantly faster and the growth of the domain causes a significant proportion of proteins to also move into the daughter bud. Thus, we expect that moving boundary effects might have important influence over previously published estimates of compartmentalisation effects in living cells, which highly depend on the ratio of fluorescence decay rates between the mother and daughter lobes [14-16].

## 3. DISCUSSION

We have numerically demonstrated that intuitive correction for moving boundaries in particle-tracking based simulators can give rise to sample paths consistent with PDEs representing the diffusion equation with moving boundaries, and that such effects can lead to non-standard effects such as super-diffusion and the creation of unusual reaction profiles. We considered both cases of particles diffusing within the boundary, and also on the boundary. While there have been previous studies showing that Master Equation approaches can incorporate growing domains [2], many drawbacks of Master Equations, such as artefacts in bimolecular reactions and boundaries, have been well-documented [5, 6]. These drawbacks motivated the approach we present here, based on particle tracking. Alternative approaches might look to extend the work of Green's Function Reaction Dynamics or First Passage Monte Carlo kinetics, but such techniques would involve numerically solving multiple moving boundary PDEs, which would cause a significant computational overhead when compared to our approach presented here.

So far, we have chosen to disregard potential advective forces arising from boundary movement. Given the dynamic nature of biological systems, with active and passive transport across membranes, it is not possible to arrive at suitable analytical forms for advective forces arising from moving boundaries in a general setting. However, if such forces could be experimentally determined, then we anticipate that our presented approach could generalise to incorporate them.

The particle-tracking algorithms for moving boundaries were benchmarked against solutions of corresponding PDEs for the case of a shrinking circle, where we considered scenarios ranging from those where the diffusive motion was slow compared to boundary movement, to those where the diffusive motion is significantly faster in comparison. For diffusion within the volume, we find that the probability distribution of quickly diffusing particles can be well-approximated by a uniform distribution at all times. The effect of the moving boundary is most noticeable at smaller diffusion coefficients, and thus we suggest that some of the most fruitful applications of moving boundary algorithms would be cases where particles diffuse slowly relative to moving boundaries. A numerical investigation into the MSD of moving particles in the static domain found that fast diffusing particles retained a profile resembling diffusion in a static domain. However, slower diffusing particles were pushed by the boundary, leading to scenarios with motion described by super-diffusion. Similar results stem from the case of diffusion on the moving boundary, where slow diffusing particles are found to show super-diffusive behaviour.

By considering the limiting case of very slow diffusion, we were able to derive analytical forms for the MSD curves corresponding to diffusion on and in the shrinking circle, and found these results to be consistent with our numerical results. We anticipate that such results should carry into higher dimensions as well, with some differences. As the integral in equation (5) shows a dependence on dimensionality, we anticipate that the degree of anomalousness of diffusion should depend on dimension. These results might motivate alternative explanations for currently documented cases of superdiffusion [17], where our observations of superdiffusion on a moving membrane might seem pertinent.

The case of bimolecular decay in an elongating dumbbell was used to demonstrate some potentially unusual effects that moving boundaries might have in non-linear systems. In particular, even though the overall volume of the dumbbell is increasing with time, we observe a bimodal distribution of reaction times. This implies that the reaction rate increases at some point in time. The reason for this is that, as the dumbbell elongates, the concentration of reactants within the bridge of the dumbbell increases, and consequently an increase in reaction rate follows. Thus, we expect that there might be reason to expect novel effects of moving boundaries in other scenarios where non-linear effects manifest themselves. Indeed, some work has already been done on investigating Turing patterns on growing domains [18], and this remains a fruitful ground for further work. Furthermore, we expect that the movement of any boundary in the direction of reactants will cause an increase in the reactant concentrations near the boundary, thus leading to an increase in reaction rates.

Finally, we concluded our sample applications by investigating photobleaching experiments in dividing *S. cerevisiae* nuclei. Such experiments have been numerically conducted on the assumption of static domains [12, 14, 16], but we demonstrate that moving boundaries can have important effects which can give rise to noticeably different results. In particular, we observed that the effective photobleaching rate for particles diffusing on nuclear membranes is higher for static boundaries than for moving boundaries, implying that existing methodologies might be underestimating photobleaching rates. Equally importantly, in formulating the model of a dividing nucleus, we were forced into making assumptions on the nature of the velocity field induced by a moving nuclear membrane. We are aware of no such empirical studies of membrane movement, so we constructed a velocity field on the basis of mapping points from the domain at one time to the closest points on the domain at another time. While mathematically convenient, it is not at all clear that budding yeast nuclear membranes specifically move in such a way, and this motivates further experiments into the movement of nuclear membranes during anaphase. It would be interesting to conduct further empirical research such that these moving boundary effects can be fully incorporated into future models. However, the current study strongly suggests that changing geometries should be included into photobleaching studies.

In conclusion, we find that intuitive based corrections to particle trackers give rise to sample paths consistent with the diffusion equation with moving boundaries. The effects of moving boundaries are diverse, ranging from causing super-diffusion in slow diffusing particles, to manipulation of local concentrations of reactants. Up to now, there has been relatively little work done in stochastic reaction-diffusion systems enclosed by moving boundaries, and it remains a fruitful ground for further research.

## Acknowledgements

We would like to thank Jai Denton, who kindly provided us with images of dividing *S. cerevisiae* nuclei for the photobleaching application. We further thank Professor Susan Gasser for kindly providing the *S. cerevisiae* strain from which the images were obtained. Lastly, we would like to thank André Leier for constructive discussion and critical reading of the manuscript.